\documentclass[a4paper,10pt,twocolumn]{article}
\usepackage{amsmath}
\usepackage{amsfonts}
\usepackage{amssymb}
\usepackage{hyperref}

\title{Comment on ``Why Einstein, Podolsky and Rosen did not prove that quantum mechanics is 'incomplete' ''}
\author{Dirk Fischer$^1$}

\begin{document}
	\twocolumn[
	\begin{minipage}{15.8cm}
	\maketitle
	\normalsize \textbf{Abstract}.
		This comment on the recently published article \textit{``Why Einstein, Podolsky and Rosen did not prove that quantum mechanics is 'incomplete' ''} (\href{http://de.arxiv.org/abs/0805.0217v1}{arXiv:quant-ph/0805.0217}) by J.H.Field shows that some conclusions, made in the referred-to article, result from an invalid use of dirac $\delta$-distributions.
		\newline
		\newline
		\newline
		\newline
	\end{minipage}
	]
\footnotetext[1]{email : dfischer@stud.physik.tu-dortmund.de}

\renewcommand{\equation}{\Roman{equation}}
\renewcommand{\theequation}{\Roman{equation}}
\section*{\normalsize Introduction}
The purpose of this comment is to point out some mathematically incorrect results in a recently published article \cite{JHField}.
Briefly, in this paper we will review the following conclusion by Field, the author of \cite{JHField} :
\begin{quotation}
``\textit{The gedanken experiment proposed by EPR cannot be carried out if the usual probabilistic  interpretation of Quantum Mechanics is correct, and so no physical conclusions can be drawn from the experi\-ment}"
\end{quotation}
and show that it is derived from an invalid calculation.

\section*{\normalsize The $\delta$-``wavefunction''}
Using the spatial representation of the well known EPR two-particle-state $|\Psi \rangle$ in the referred-to paper's eq.(5), that is $\langle x_1,x_2|\Psi \rangle = \Psi(x_1,x_2) = h\ \delta(x_1-x_2+x_0)$, where the $x_i$, $i=1,2$, denote the coordinates of particle 1 and 2, respectively, and $x_0$ their origin, Field gives an expression for the probability $P(a \leq x_1 \leq b)$ of finding particle 1 in an interval $[a,b]$ with $a,b \in \mathbb{R}$ and $a \leq b$ while particle 2 might have any arbitrary position possible. Calculating this probabilty, usage is made of ``\textit{the usual probabilistic interpretation of Quantum Mechanics}" by writing
\begin{eqnarray}
 & & P(a \leq x_1  \leq b) \nonumber \\
 & = & \lim_{L\to \infty} \ \frac{\int_a^b \mathrm{dx}_1 \int_{-\infty}^\infty \mathrm{dx}_2 |\Psi(x_1, x_2)|^2}{\int_{-L}^L \mathrm{dx}_1 \int_{-\infty}^\infty \mathrm{dx}_2 |\Psi(x_1, x_2)|^2} \label{EQ1}\\
 & = & \lim_{L \to \infty} \ \frac{b-a}{2L}, \nonumber
\end{eqnarray}
rendering eq.(6) in Field's paper.
As Field correctly remarks, the above given wavefunctions (in the nominator or denominator) are not square-integrable (taking the limit $L\to \infty$ for the $\mathrm{dx_1}$-integration), leading to the referred-to paper's eq.(6) and (8) with $P(a \leq x_1  \leq b) \to 0$. This vanishing lets the author of the referred-to paper correctly dismiss the usage of $\Psi(x_1,x_2) = h\ \delta(x_1-x_2+x_0)$ for purposes of expressing eq.(\ref{EQ1}).
\newline
Within the following lines a new aspect has to be investigated - the fact that not eq.(\ref{EQ1}) vanishes in the limit $L \to \infty$, as it is stated in the referred-to paper's eq.(6), but that it is not mathematically well-defined at all. It is the use of $\delta$-distributions in the integral's kernel, $|\Psi(x_1, x_2)|^2$, that leads to difficulties and violation of the quantum mechanical requirements for proper wavefunctions by the dirac-$\delta$. As a consequence, not only ``\textit{no meaningful conclusions can  be drawn from any gedanken experiment based upon non square-integrable wave functions}'' as it is written in the referred-to paper, but not even mathematically meaningful expressions can be gained from such functions. It will be shown that Field's correct criticism on non-square-integrable wavefunctions applies to methods used in his paper, too.
After dealing with certain preliminaries in this section, in the section afterwards light is shed on the referred-to paper's main proof.
\newline
\newline
The question why the probabilty-interpretation for eq.(\ref{EQ1}) completely fails, is answered only in part by the property ``non-square-integrabilty'' (as it is usually called if an integral similar to the denominator of (\ref{EQ1}) has a definite value $< \infty$  before taking the limit of the boundaries $\to \pm \infty$, but then diverges during that limit). At least in the same manner severe complications arise  from the distributional character of the intergral's kernel - even before taking the limit.
\newline
In our case the (norm-) square of $\Psi(x_1, x_2)$ contains an undefined product of dirac-$\delta$, i.e. $\delta^2$. Considering the common theory of distributions by Sobolev and Schwartz, such a multiplication of distributions is not defined in general within the vectorspace of distributions $\mathcal{D}'$. It was shown in 1954 by Schwartz \cite{Schwartz-proof}, that $\mathcal{D}'$ even couldn't be embedded into into an associative Algebra $(\mathcal{A},+,\circ)$ (with certain special features that are omitted here for brevity) which could have comprised a consistent multiplication of distributions.
It lasted until 1984 to create a theory which succeeded in providing the framework for a slightly altered but general multiplication of distributions, the so called Colombeau version of distribution theory \cite{Colomb1}. Within this theory it can be shown that :
\begin{enumerate}
 \item $\delta^2$ hasn't any counterpart in the disitribution theory of Sobolev and Schwartz (i.e. $\delta^2$ isn't associated with any $\iota(\omega)$, $\omega \in \mathcal{D}'$ where $\iota$ is the embedding of $\mathcal{D}'$ into Colombeau's associative algebra) and
 \item $\delta^2$  in the Colombeau version, applied to any testfunction, leads to a divergent expression (of the order $1/\epsilon$, with $\epsilon \to 0$).
 Therefore even the term $\int_{-\infty}^{\infty} \mathrm{dx} \ \delta^2(x)$ itself is divergent and then especially $\int_{-\infty}^{\infty} \mathrm{dx_1} \int_{-\infty}^{\infty} \mathrm{dx_2} \ \delta^2(x_1-x_2)$.
\end{enumerate}
This leads to the inconvenience of eq.(\ref{EQ1}) not being defined, even before taking the limit $L \to \infty$ because 
\begin{eqnarray*}
& & \int_a^b \mathrm{dx}_1 \int_{-\infty}^{\infty} \mathrm{dx}_2 \ |\Psi(x_1, x_2)|^2 \\
& = & \int_a^b \mathrm{dx}_1 \int_{-\infty}^{\infty} \mathrm{dx}_2 \ \delta^2(x_1 - x_2 + x_0) \\
& = & \int_a^b \mathrm{dx}_1 \cdot \infty ,\\
\end{eqnarray*}
which is not defined, resulting in
\begin{eqnarray*}
 (\mathrm{I}) & = & \lim_{L \to \infty} \frac{\infty \cdot (b-a)}{\infty \cdot 2L}, \label{infty}
\end{eqnarray*}
again obiously not being defined.
Formally one could write, as it is in perturbative field theo\-retic approaches often done, $\int_a^b \mathrm{dx} \ \delta^2(x) = \delta(0)$ and then
\begin{eqnarray}
 & = & \int_a^b \mathrm{dx}_1 \int_{-\infty}^{\infty} \mathrm{dx}_2 \ \delta^2(x_1 - x_2 + x_0) \nonumber \\
 & = & \delta(0) \int_a^b \mathrm{dx}_1 \label{EQ3} \\
 & = & (b-a) \cdot \delta(0), \nonumber
\end{eqnarray}
with $\delta(0)$ being a ``divergent factor", leading to
\begin{eqnarray*}
 (\mathrm{I}) & = & \lim_{L \to \infty} \frac{\delta(0) \cdot (b-a)}{\delta(0) \cdot 2L}.
\end{eqnarray*}
Within Colombeau's theory it is possible to show the divergence of the disitribution $\delta(x)$ for $x=0$. This enables us to (roughly) identify the above mentioned formal factor $\delta(0)$ with $\delta(x)|_{x=0}$, ma\-king the formal equation (\ref{EQ3}) more plausible and stressing out $\delta(x)|_{x=0} \neq 1$ - as it is often wrongly taken.
\newline
\newline
We conclude that not ``\textit{the 'relative probabilty' $P(a,b)$ of EPR's Equation $(6)$ also vanishes}" as it is stated in the referred-to paper, but the relative probabilty is not well-defined at all if a $\delta$-disitribution is taken as ``wavefunction". So it is invalid using eq.(\ref{EQ1}) for any considerations regarding $P(a,b)$.
\newline
Of course one might try to ``rescue" eq.(\ref{EQ1}) by normizing $\Psi(x_1,x_2)$ with a factor of $1/\sqrt{\delta(0)}$. But this formally gives us, using the known rules for manipulations of the dirac-$\delta$ :
\newline
$\Psi(x) = \delta(x)/\sqrt{\delta(0)} = \delta(x \cdot \sqrt{\delta(0)})$, leaving opportunity for interpretation; additionally, taking the limit $L \to \infty$ again would rise a problem.
\section*{\normalsize ``Minimally modified'' wavefunction}
Nearly all results, found in the preceeding section, object to Field's approach of calculating $P(a \leq x_1 \leq b)$ with a "minimally modified" wavefunction (eq.(9) in his paper), given by
\begin{eqnarray}
 & & \tilde{\Psi}(x_1, x_2) \nonumber\\
 & = & \frac{1}{(\sqrt{2\pi}\sigma_x)^{1/2}} \cdot \exp \left( \frac{x_0^2 - 2x_1^2 - 2x_2^2}{16 \sigma_x^2} \right) \nonumber \\
 &  & \cdot \delta(x_1 - x_2 + x_0). \label{EQ2}
\end{eqnarray}
Contrary to the claim in the referred-to paper (eq.(10) in that paper), this $\tilde{\Psi}(x_1, x_2)$ again lacks of the wavefunction's requirements and is not square-integrable - due to the use of $\delta$-distributions. With (\ref{EQ2}) we get
\begin{eqnarray*}
& & P(a \leq x_1 \leq b) \\
& = & \int_a^b \mathrm{dx}_1 \int_{-\infty}^{\infty} \mathrm{dx}_2 \ |\tilde{\Psi}(x_1, x_2)|^2 \\
& = & \int_a^b \mathrm{dx}_1 \int_{-\infty}^{\infty} \mathrm{dx}_2 \ \delta^2(x_1-x_2+x_0) \ f^2(x_1,x_2),
\end{eqnarray*}
where $f(x_1,x_2)=1/(\sqrt{2\pi}\sigma_x^2) \cdot \exp([x_0^2-2x_1^2-2x_2^2])/16\sigma_x^2$ as it was given in eq.(\ref{EQ2}).
Performing the first integration, this yields (adopting the notion of a``divergent factor'' $\delta(0)$), similarly to the case in eq.(\ref{EQ3}),
\begin{eqnarray*}
 & & \int_a^b \mathrm{dx}_1 \int_{-\infty}^{\infty} \mathrm{dx}_2 \ \delta^2(x_1-x_2+x_0) \ f^2(x_1,x_2) \\
 & = & \delta(0) \int_a^b \mathrm{dx}_1 \ f^2(x_1=x_2-x_0)
\end{eqnarray*}
again showing the formal divergence $\delta(0)$.
\newline
Cancelling the dirac-$\delta$ from eq.(\ref{EQ2}) would solve the mathematical problems but at the same time jeopardize Field's further reasoning where the constraint $x_2 = x_1 + x_0$ of the mutual particle positions, ensured by the $\delta$, plays a crucial role.
Without this constraint the argumentation used in the referred-to paper, namely ``\textit{measuring $x_1$ in the interval $\delta x_1$ then enables the certain prediction that $x_2$ lies in the interval $\delta x_2$ around $x_2 = x_1 + x_0$}" doesn't hold anymore. This is because without a $\delta(x_1-x_2+x_0)$, the two-particle wavefunction allows for particle positions $x_1$ and $x_2$ with $x_2 \neq x_1+x_0$ and non-vanishing probabilty. In other words, the vanishing probability of one particle being found in an ever decreasing interval $\delta x_i$ doesn't affect the other particle's position-probabilty.
\newline
In fact, the $\tilde{\Psi}$ without the $\delta$, let us call it $\Psi'$ with
\[
 \Psi'(x_1, x_2) = \frac{1}{(\sqrt{2\pi}\sigma_x)^{1/2}} \cdot \exp \left( \frac{x_0^2 - 2x_1^2 - 2x_2^2}{16 \sigma_x^2} \right),
\]
isn't anymore a suitable wavefunction for describing a sharply position-entangled EPR-pair of particles.
\section*{\normalsize Discussion}
The preceeding sections do not provide a proof to the contrary of the fundamental concerns raised by J.H.Field, but show that the proof of the referred-to paper's main claim needs some additional work.
\newline
Another important question, concerning the foundations of a mathematically consistent theory, became obvious throughout the above lines : If the $\delta$, being used as a position-eigenvector in the spatial-representation (i.e. $\langle x | x' \rangle = \delta(x-x')$), is divergent for $x=x'$ - then to what extent should it be used as eigenvector in quantum mechanics ?
As Field correctly remarks, the dirac-$\delta$ is ``\textit{a mathematical idealisation never realized in the wavefunction of any actual physical system}'', hinting to the emerging problems (see e.g. reference [7] in the referred-to paper). In fact, we saw it isn't a mathematically valid wavefunction (i.e., loosely speaking, an element of the hilbertspace of the square-integrable functions) at all.

\end{document}